# Solid angles in perspective

Paul Quincey

National Physical Laboratory, Hampton Road, Teddington, TW11 0LW, United Kingdom
E-mail: paulgquincey@gmail.com

**Abstract**

The specialised uses of solid angles mean that they are quite unfamiliar quantities.  This article, apart from making solid angles a little more familiar, brings out several topics of general interest, such as how units are interrelated and how equations depend on the choice of units. Although the steradian is commonly used as the unit for solid angle, another unit, the square degree, is used in astronomy, and a unit introduced here, the solid degree (with 360 solid degrees in a hemisphere) could be used with benefits that are similar to those of the degree when it is used as the unit for plane angle. The article, which is suitable for students at A-level and introductory undergraduate level, also shows how solid angles can provide a gentle introduction to crystal structure, spherical trigonometry and non-Euclidean geometry.

**Introduction**

Solid angles, usually given the symbol $\Omega$, have a small but essential role in physics. For example, how can you characterise the intensity of a bicycle light? Its optical power, the total energy of the photons emitted per second, is clearly important, but this does not take account of the spread of the beam, which will determine the intensity in a particular direction. What is needed is a measure of the "angular area" of the light beam. The spread in one direction is measured as a plane angle; the spread over an area, which is what we are interested in here, is measured as a solid angle.

The power of the light is called radiant flux, measured in watts (in SI units). The SI unit for solid angle is the steradian (sr), and the radiant intensity is measured in watts per steradian. When the sensitivity of a typical human eye is taken into account, the power is called luminous flux, expressed in lumens, and the luminous intensity is given in lumens per steradian, otherwise called candelas. The increasing dazzle-power of modern bicycle lights is likely to lead to regulation limiting their luminous intensity to 200 candelas. The field of view of a human eye is around 4 sr, so the steradian is a convenient size in this context.

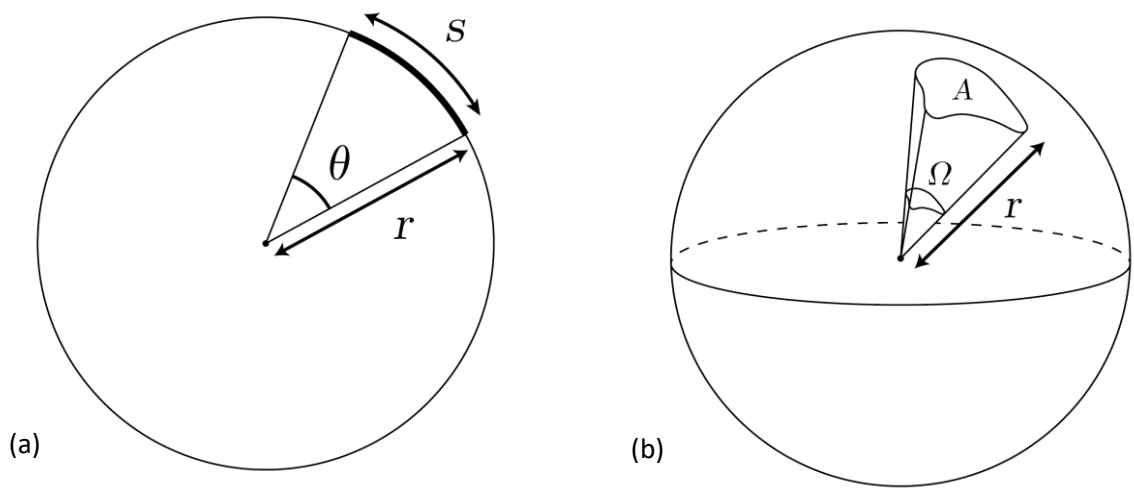

(a)        (b)

Figure 1: (a) a plane angle $\theta$ at the centre of a circle of radius $r$, with the arclength $s$ that it projects on the circumference, and (b) a solid angle $\Omega$ at the centre of a sphere of radius $r$, with the area $A$ that it projects on the surface of the sphere.

Solid angles are also needed when astronomers want to quantify the area of a patch of sky. They often use square degrees ($deg^2$). A single square degree is easily visualised as a small square patch of sky, where the view through a telescope would move from one corner of the square to an adjacent corner by rotating the telescope one degree. The sun and moon both appear as circular areas of sky, with solid angles, as seen from the Earth, of about 0.2 $deg^2$. This would be about 60 microsteradians, so in this case the square degree is a more convenient size than the steradian.

**Complete equations**

As discussed in Quincey and Burrows (2019), equations that are valid for any coherent set of units are called "complete" or unit-invariant equations. This type of equation must be used if deductions are to be made about the dimensions of the quantities in the equation. If an equation is valid only for a specific unit, or only one set of coherent units, it is not complete. It is instead "unit-specific", and it cannot be used to deduce dimensional relationships.

As an example, the familiar equation linking a plane angle $\theta$ to an arclength $s$ (as shown in Figure 1a) is $\theta = s/r$, but this is not complete, because it is only correct when the unit used for the angle is the radian. The less familiar complete equation is:

$$\theta = \theta_N \, s/r \qquad (1)$$

where $\theta_N$, called the Cotes angle, is a constant, an angle measuring 1 rad or $180/\pi$ degrees, or the same quantity in whatever angle unit is being used[1].

**The definition of solid angle $\Omega$**

Solid angles of equal size can be very different shapes, in the same way that equal areas can be very different shapes. It is important to note that the solid angle is located in the vicinity of where the radiating lines meet, and it has a well-defined value whether or not a sphere is somehow drawn around it. In the same way, the angle between two stars as seen from Earth is well-defined, even though their distances from Earth may be unknown, making the circle in Figure 1a entirely notional.

The size of a solid angle is, however, most simply defined in terms of the area of spherical surface that a solid angle would project onto a sphere centred on its apex, as shown in Figure 1b. This definition is usually presented as $\Omega = A/r^2$. The area and radius can be in any coherent units, but the value for $\Omega$ given by this equation is always in steradians. If we call the solid angle of a full sphere $\Omega_{sph}$, this equation gives the value of $\Omega_{sph}$ to be $4\pi$, which is only correct when the unit is the steradian, so the equation is not complete.

If square degrees are used, the definition of $\Omega$ becomes $\Omega = 180^2/\pi^2 \, A/r^2$, another non-complete equation. The value of $\Omega_{sph}$ in square degrees is $360^2/\pi \approx 41{,}253$.

Of course the equation in steradians is particularly simple, but the complete equation relating area to solid angle needs to include the Cotes angle explicitly, analogously to the equation for arclength (1):

$$\Omega = \theta_N^2 A/r^2 \qquad (2)$$

We can call $\theta_N^2$, which is equal to one steradian, the Cotes solid angle. The dimensional relationship is that solid angles have the dimensions of $(angle)^2$. When using the simplified equations that require the use of radians and steradians, in effect we treat the radian as a natural unit and set $\theta_N$ equal to 1 (Quincey, 2016). The act of implicitly setting $\theta_N$ equal to 1 has been called "the Radian Convention" (Quincey and Burrows, 2019).

---

[1] The constant $\theta_N$ has appeared in other published work as various different symbols: 1/□ (Brownstein, 1997), 1/$\eta$ (Quincey, 2016), or simply rad (various). Although $\theta_N$ = 1 rad, it is much clearer to give the constant and the unit different symbols. The N signifies that the radian is the natural unit of angle.

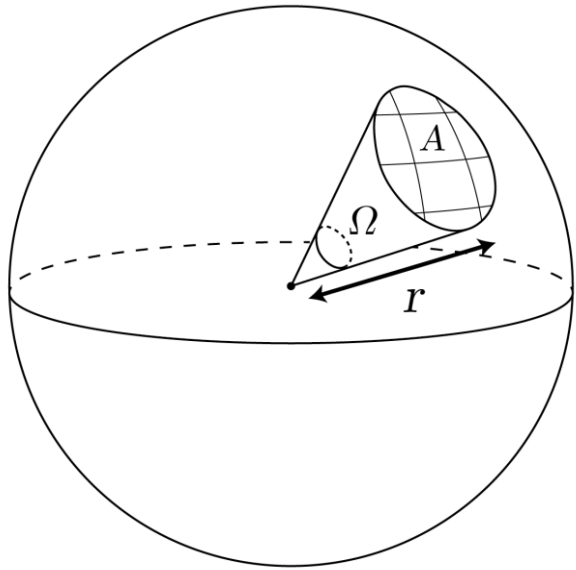

Figure 2: A conical solid angle Ω at the centre of a sphere of radius *r*, with the area *A* that it projects on the surface of the sphere. The solid angle is completely determined by the cone's apex angle, the angle subtended at the centre of the sphere by the diameter of its circular cross-section.

**Cones**

We tend to think that the area *A* within a circle of radius *r* is always given by $A = \pi r^2$, whatever its size, but this is only true when the surface enclosed by the circle is flat. For a curved surface like the surface of a sphere, the equation will only hold when the circle is small compared to the radius of curvature. Similarly, the solid angle Ω of a cone with an apex angle of 2*θ* is only equal to $\pi\theta^2$ in the limit of small angles.

For a sphere of radius *r*, the area of its surface marked out by the conical solid angle can be calculated using basic calculus to be[2] $2\pi r^2 (1 - \cos\theta)$. From Equation 2, then, the solid angle of a cone is given by the complete equation:

$$\Omega = 2\pi\theta_N^2 (1 - \text{Cos } \theta) \qquad (3)$$

We can see that in the limit of small *θ*, we have[3] $\Omega = \pi\theta^2$. An angular radius *θ* of 1° gives a solid angle $\Omega = 0.99997\,\pi\theta^2$, showing that solid angles follow the rules for flat surfaces very closely at that scale. In contrast, with an angular radius *θ* of 1 rad, $\Omega = 0.9194\,\pi\theta^2$.

**Square pyramids**

The solid angles found in the natural world, such as in crystal structures, are generally bounded by planes and so are not cones. The mathematics becomes more complicated, but it also provides a gentle introduction to spherical trigonometry. The simplest case is a square pyramid.

How do you calculate the area of the "square" patch that the square pyramid of Figure 3 would project onto the surface of a surrounding sphere? Fortunately, the answer follows directly from a theorem in spherical trigonometry called the Spherical Excess theorem.

---

[2] Trigonometric functions like sin *θ* and cos *θ* need more care when angle units other than radians are allowed. The functions are commonly used in two distinct ways: sin x, where x is dimensionless, and the familiar expansion $\sin x = x - \tfrac{1}{6}x^3 + \ldots$ applies, and sin *θ*, where *θ* is an angle (in any angle unit), and the expansion is $\sin\theta = \theta/\theta_N - \tfrac{1}{6}(\theta/\theta_N)^3 + \ldots$. Brownstein (1997) suggested using capital letters to distinguish the two: sin x and Sin *θ*, so that $\text{Sin } \theta = \sin(\theta/\theta_N)$. This paper follows that convention.

[3] Following the Brownstein convention, in the limit of a small angle *θ*, $\text{Cos } \theta = 1 - \tfrac{1}{2}(\theta/\theta_N)^2$.

If β is the angle between the faces of the pyramid (see Figure 3), the Spherical Excess theorem, as normally written, states that the solid angle Ω at the apex of the pyramid is given by Ω = 4β – 2π. This is an extraordinarily simple result[4]. It appears to mix solid angles (Ω), plane angles (β) and numbers (π), as if they are all dimensionless, but again the equation is unit-specific. Ω must be in steradians, and β in radians, for it to be valid, so we must avoid the temptation to infer dimensions from the equation in this form.

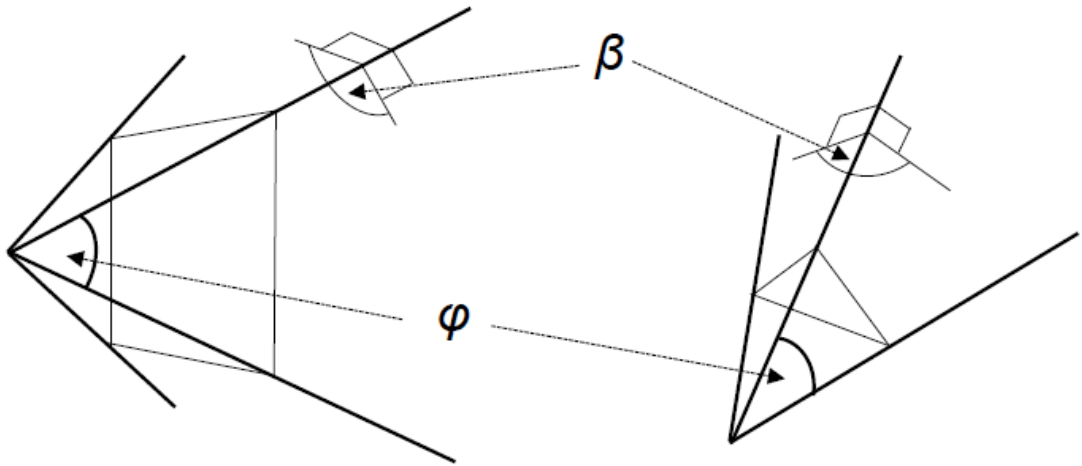

Figure 3: Square pyramids and equilateral triangular pyramids.

The complete equation for the solid angle at the apex of a square pyramid is:

$$\Omega = 4\theta_N \beta - 2\pi \theta_N^2 \qquad (4)$$

where any coherent units for angle and solid angle can be used. As before, the familiar equation is obtained by setting $\theta_N$ equal to 1, but this is done purely to simplify the equation, and it removes the information about dimensions.

The angle between faces (β) is not very convenient, but we can use trigonometry and vectors to relate this to other angles, such as the angle at the vertex of each face, here called φ (see Figure 3).

It can be shown that Cos β = - Tan² ½φ, and if we use this to replace β with φ in (4), we have

$$\cos \beta = \cos(\tfrac{1}{4} \Omega/\theta_N + \tfrac{1}{2} \pi \theta_N) = -\sin \tfrac{1}{4} \Omega/\theta_N$$

and so $\qquad \sin \tfrac{1}{4}\Omega/\theta_N = \tan^2 \tfrac{1}{2}\varphi. \qquad (5)$

When the Radian Convention, to set $\theta_N$ equal to 1, is adopted, this becomes the slightly simpler equation sin ¼Ω = tan² ½φ, which raises the awkward question of what the sine of a solid angle represents, apart from the abstract mathematical function. We can see from the complete equation that before the equation is simplified, the sine function is actually being applied to a quantity with the dimensions of plane angle.

In the limit of small angles, equation (5) becomes Ω = φ², analogous to the area of a square. The value of the face vertex angle φ can vary from 0, for a very thin pyramid, to 90° or π/2 rad for a flat one. Equation (5) tells us that Ω is 0 and 2π sr for these angles respectively, as we would expect.

Curiously, perhaps the most familiar square pyramid in the world, at Cheops in Egypt, has an apex solid angle very close to π/2 sr.

---

[4] This is also known as Girard's theorem, after Albert Girard who published it in 1626. The basic theorem, for triangles, can be seen to be true by drawing three great circles on a convenient sphere such as a table tennis ball.

**Equilateral triangular pyramids**

The other shape we will consider here is a pyramid whose base is an equilateral triangle (see Figure 3). Applying the Spherical Excess theorem again, the complete equation for the solid angle of the apex of the pyramid is:

$$\Omega = 3\theta_N \beta - \pi \theta_N^2 \qquad (6)$$

Again the relationship between $\beta$ and $\varphi$ can be deduced from trigonometry and vectors, in this case being $2\cos \beta = 1 - \tan^2 \tfrac{1}{2}\varphi$.

So we have the slightly more complicated relationship between $\Omega$ and $\varphi$ of:

$$2\cos \tfrac{1}{3}(\Omega/\theta_N + \pi\theta_N) = 1 - \tan^2 \tfrac{1}{2}\varphi \qquad (7)$$

This time, the value of the face vertex angle $\varphi$ can vary from 0, for a very thin pyramid, to 120° or $\pi/3$ rad for a flat one. Equation (7) tells us that $\Omega$ is 0 and $2\pi$ sr for these angles respectively, again as we would expect. In the limit of small angles, Equation (7) becomes $\Omega = \sqrt{3}\,\varphi^2/4$, analogous to the area of an equilateral triangle.

**Introducing the solid degree**

The great benefits of degrees, rather than radians, as units for angle are that simple fractions of a full rotation give us simple whole numbers of degrees: 30°, 45°, 120° and so on, and, moreover, when an angle isn't a simple fraction of a rotation, like 46.4°, it is very easy to compare this with the numbers that are simple fractions. It is not readily apparent if an angle of 0.535 rad is more or less than $\pi/6$ rad, for example. The familiarity of degrees also means that relations such as Sin 30° = 0.5 and Tan 45° = 1 are well known.

You might think that the square degree unit for solid angle would also have these benefits, but it does not. The solid angle of a cube corner (or the apex of great pyramid of Cheops), $\pi/2$ sr, is roughly 5157 deg$^2$.

Of course, in physics we can choose to use any units we like, for convenience, providing we understand what we are doing. The important questions are whether the units are coherent with the other units we are using, and whether the equations we are using are complete (correct for any coherent set of units), or unit-specific (where more care is required).

We could divide the solid angle of a sphere, $\Omega_{sph}$, into 360 new units, by analogy with degrees, but it is more useful to divide $\Omega_{sph}$ into 720 equal units, which we will call solid degrees, abbreviated as sd. These have the simple property that $2\pi$ sr = 360 sd, the same relationship as $2\pi$ rad = 360°. The solid angle of a cube corner (or the apex of the great pyramid) is 90 sd. A solid angle of 1 sd is roughly the size of the apex of a square pyramid with a face angle $\varphi$ of 7.6° - about half the solid angle at the top of the Shard building in London.

We met trigonometric functions involving solid angles above. One useful feature of solid degrees is that Sin $\Omega/\theta_N$, where $\Omega$ is a solid angle in solid degrees, equals Sin $\Omega$, where $\Omega$ is a plane angle in degrees[5], the same convenient relationship as with steradians and radians. Relations such as Sin 30/$\theta_N$ = 0.5, where the 30 represents 30 sd, therefore appear familiar. Also, the equations for the solid angles of square and triangular pyramids, (4) and (6), when made unit-specific by having $\beta$ in degrees and $\Omega$ in solid degrees, are simply $\Omega = 4\beta - 360$ and $\Omega = 3\beta - 180$ respectively.

You may be wondering how the degree unit for plane angle can give rise to two different units for solid angle. The answer is that while the square degree and the degree are coherent units, the solid degree and the degree are not. In fact the unit for plane angle that is coherent with the solid degree has a value of $\sqrt{(180/\pi)}° \approx 7.57°$. This could be called a root solid degree (rsd), The solid degree, then, is a convenient unit for solid angle, and makes a good case study for demonstrating how units and equations are interrelated, but it is a bad candidate for a coherent unit system.

---

[5] For a solid angle $\Omega$, Sin $\Omega/\theta_N$ = sin $\Omega/\theta_N^2$ where $\theta_N^2$ is 180/π solid degrees, as set out in Table 1.
For a plane angle $\Omega$, Sin $\Omega$ = sin $\Omega/\theta_N$ where $\theta_N$ is 180/π degrees, as set out in Table 1

The basic facts about the three solid angle units featured above are given in Table 1.

| Solid angle unit | Spherical solid angle, $\Omega_{sph}$ | Cotes solid angle, $\theta_N^2$ = $\Omega_{sph}/4\pi$ | Unit in terms of the other units | Coherent unit for plane angle |
|---|---|---|---|---|
| Steradian (sr) | $4\pi$ sr ≈ 12.57 sr | 1 sr | 1 sr = $180^2/\pi^2$ deg$^2$ ≈ 3283 deg$^2$<br>1 sr = $180/\pi$ sd ≈ 57.3 sd | Radian (rad)<br>= $180/\pi$ deg ≈ 57.3° |
| Square degree (deg$^2$) | $360^2/\pi$ deg$^2$ ≈ 41,253 deg$^2$ | $180^2/\pi^2$ deg$^2$ ≈ 3,283 deg$^2$ | 1 deg$^2$ = $\pi/180$ sd ≈ 0.0175 sd<br>1 deg$^2$ = $\pi^2/180^2$ sr ≈ 3.05x10$^{-4}$ sr | Degree (deg or °) |
| Solid degree (sd) | 720 sd | $180/\pi$ sd ≈ 57.3 sd | 1 sd = $180/\pi$ deg$^2$ ≈ 57.3 deg$^2$<br>1 sd = $\pi/180$ sr ≈ 0.0175 sr | Root solid degree (rsd)<br>= $\sqrt{(180/\pi)}$ deg ≈ 7.57° |

Table 1: a selection of solid angle units and their coherent plane angle units

**Some naturally occurring plane and solid angles**

Like the radian, the steradian has mathematical rather than physical significance. We do not expect to find angles in the natural world measuring exactly 1 radian. What we often find are simple fractions of 360° or $2\pi$ rad – but not always. Similarly, solid angles in the natural world are usually simple fractions of 720 sd or $4\pi$ sr.

Any collection of plane angles that add to 360° can be assembled to form the centre of a disc, but a collection of solid angles that add to 720 sd forms a three-dimensional puzzle, and will not necessarily fit together to form the centre of a sphere. The solid angles around any atom within a crystal structure must necessarily add to 720 sd, making them interesting practical examples. Incidentally, the availability of 3D printing makes the task of illustrating how solid angles fit together much easier than in the past.

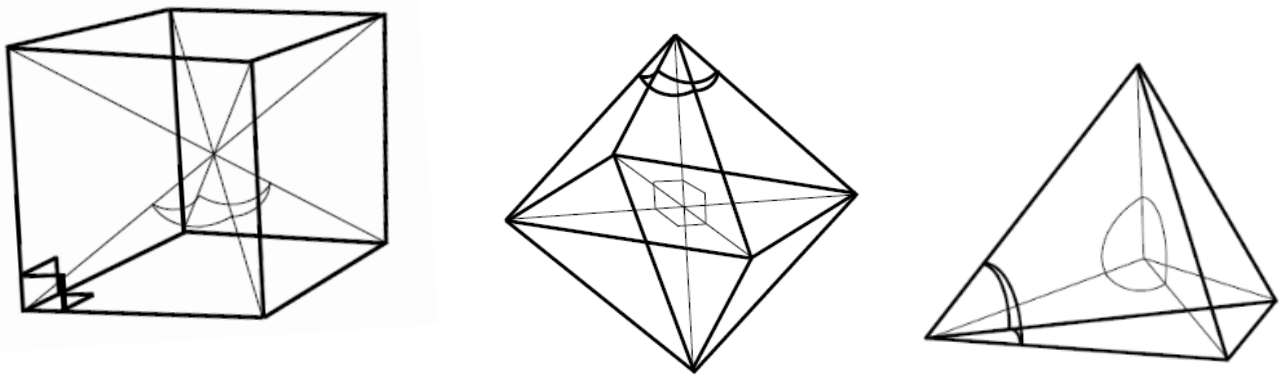

Figure 4: the solid angles at the corners (thick lines) and centres (thin lines) of a cube, octahedron and tetrahedron

The most basic naturally-occurring angles and solid angles can be found within the cube, octahedron and tetrahedron, as illustrated in Figure 4. The solid angles of their corners, and at their centres (shown in Figure 4), in steradians and solid degrees, together with the associated plane angles (as shown in Figure 3), in radians and degrees, are given in Table 2.

|  | Square or Triangular | Solid angle $\Omega$ (sr) | $\Omega$ (sd) | Face vertex angle $\varphi$ (rad) | $\varphi$ (deg) | Angle between faces $\beta$ (rad) | $\beta$ (deg) |
|---|---|---|---|---|---|---|---|
| Cube corner | T | 1.571 = π/2 | 90 | 1.571 = π/2 | 90 | 1.571 = π/2 | 90 |
| Cube centre | S | 2.094 = 2π/3 | 120 | **1.231** | **70.53** | 2.094 = 2π/3 | 120 |
| Octahedron corner | S | **1.359** | **77.88** | 1.047 = π/3 | 60 | **1.911** | **109.47** |
| Octahedron centre | T | 1.571 = π/2 | 90 | 1.571 = π/2 | 90 | 1.571 = π/2 | 90 |
| Tetrahedron corner | T | **0.5513** | **31.59** | 1.047 = π/3 | 60 | **1.231** | **70.53** |
| Tetrahedron centre | T | 3.142 = π | 180 | **1.911** | **109.47** | 2.094 = 2π/3 | 120 |

Table 2: some naturally-occurring angles and solid angles in cubes, octahedrons and tetrahedrons. The ones that are not simple fractions of a circle or sphere are in bold.

Along with the simple fractions of 2π rad (360°) and 4π sr (720 sd), these simple shapes provide some other naturally-occurring plane and solid angles, highlighted in bold in Table 2. These are[6]:

| | | |
|---|---|---|
| 1.231 rad | = 70.53 deg | = arcCos 1/3 |
| 1.911 rad | = 109.47 deg | = arcCos -1/3 |
| | | |
| 0.5513 sr | = 31.59 sd | = $(3\theta_N \text{ arcCos } 1/3) - \pi\theta_N^2 = \theta_N \text{ arcCos } 23/27$ |
| 1.359 sr | = 77.88 sd | = $4\theta_N \text{ arcSin } 1/3$ |

It is a curious geometrical fact that 8 tetrahedron corners (0.5513 sr each) and 6 octahedron corners (1.359 sr each) make a total solid angle of exactly 4π sr that will fit together to form the centre of a sphere. The crystal structure formed like this is found in diamond, alum and fluorite, and is sometimes called a tetrahedral octahedral honeycomb, as shown in Figure 5.

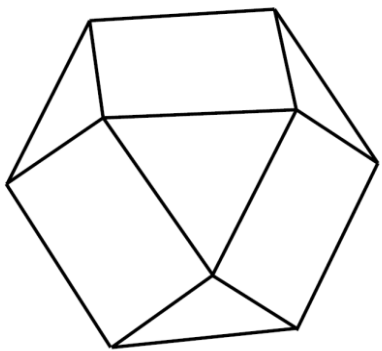

Figure 5: the crystal structure formed by 6 octahedrons and 8 tetrahedrons

---

[6] As we have seen, in the Brownstein convention y = cos x turns the number x into the number y, while b = Cos a turns an angle a into a number b. In the same way, arccos y turns a number y back into the number x, or rather a set of possible numbers including x, while arcCos b turns the number b into an angle a, or rather a set of possible angles including a, whose units need to be specified. ArcSin and arcCos can therefore give results in rad, deg, or any other angle unit, as a free choice, with the unit-appropriate value of $\theta_N$, such as those in Table 1.

**Discussion and summary**

Solid angles are directly encountered quite rarely, and they have rather specialised uses, such as quantifying the intensity of a light source. However, the mathematical treatment of spherical surfaces is relevant to many areas of physics. The appearance of 4π in the equation for Coulomb's law (for the force between two charges), to avoid 4π appearing in the equation for Gauss's Law (for the electric field around a charge), is a good example. Familiarity with the concepts and equations for solid angles can only be helpful for understanding these topics.

The excursion into different units for plane and solid angle could be seen as a good reason for picking one set and sticking to it. In mathematics, it is conventional not just to always use radians and steradians, but also to "build in" this convention to the equations that are used, leading to simplified unit-specific equations, like $\Omega = A/r^2$. If other units are used, it is tempting to say that "the equations don't work", and therefore that using other units is wrong. The point is rather that "the *familiar* equations don't work". We should always be aware if the equations we are using are unit-specific, and remember that unit-invariant (or complete) versions of them can always be made by inserting one or more suitable constants. Only complete equations give the true dimensional relationships between quantities.

Familiarity with equations, which may be simplified, is not necessarily the same thing as familiarity with the relations between physical quantities. It is all very well to say "if you always use radians and steradians, you don't have to think about them," but we should all think about what we are doing from time to time.

**Acknowledgements**

I would like to thank Kathryn Burrows, Andrew Lewis and Tom Quincey for very helpful comments on the manuscript. Tom Quincey also prepared some of the diagrams.